# Progressive Transfer Learning for Multi-Pass Fundus Image Restoration

Uyen Phan, Ozer Can Devecioglu, Serkan Kiranyaz and Moncef Gabbouj, IEEE Fellow.

***Abstract*—** Diabetic retinopathy (DR) is a leading cause of vision impairment, making its early diagnosis through fundus imaging critical for effective treatment planning. However, the presence of poor-quality fundus images—caused by factors such as inadequate illumination, noise, blurring and other motion artifacts yields a significant challenge for accurate DR screening. In this study, we propose progressive transfer learning (PTL) for multi-pass restoration to iteratively enhance the quality of degraded fundus images, ensuring more reliable DR screening. Unlike previous methods that often focus on a single-pass restoration, multi-pass restoration via PTL can achieve a superior blind restoration performance that can even improve most of the good-quality fundus images in the dataset. Initially, a Cycle-GAN model is trained to restore low-quality images, followed by PTL induced restoration passes over the latest restored outputs to improve overall quality in each pass. The proposed method can learn blind restoration without requiring any paired data while surpassing its limitations by leveraging progressive learning and fine-tuning strategies to minimize distortions and preserve critical retinal features. To evaluate PTL's effectiveness on multi-pass restoration, we conducted experiments on DeepDRiD, a large-scale fundus imaging dataset specifically curated for diabetic retinopathy detection. Our result demonstrates state-of-the-art performance, showcasing PTL's potential as a superior approach to iterative image quality restoration.

## I. Introduction

Diabetic retinopathy (DR) is an important eye condition caused by blocked blood vessels in the retina. In response, the body grows new vessels, which can lead to retinal detachment and blindness. Projections estimate that by 2030, approximately 191 million people worldwide will be affected by diabetic retinopathy (DR), with around 56.3 million experiencing vision-threatening stages of the disease [1]. Detecting DR in its early stages is crucial for effective treatment. However, the disease often presents no noticeable symptoms in its initial phases, apart from minor visual impairments. Given the subtlety of these early signs, the World Health Organization (WHO) stresses the vital need for regular screenings to detect the condition before it progresses unnoticed [2]. Leveraging advanced diagnostic techniques like fundus imaging has become an increasingly critical tool to aid early detection for DR. This technique provides detailed visualizations of the retina's internal structure, enabling the identification of anatomical abnormalities linked to DR.

Recent advancements in machine learning, spanning diverse modalities, have further enhanced retinal diagnostics. By integrating these technologies, digital fundus images now support automated diagnosis modules. Various approaches have been proposed for automatic DR detection systems in literature, spanning from traditional machine learning to deep learning. Traditional methods focus on isolating salient lesions such as microaneurysms, blood vessels, hemorrhages, and exudates. For instance, techniques such as morphological operations on fundus images combined with CLAHE [3] (Contrast Limited Adaptive Histogram Equalization) enhance the visibility of anomalies. Other methods involve image transformation methods like wavelet, curvelet, and contourlet to enhance the shape of retinal components [4]-[7]. Many of these approaches concentrate on detecting specific lesions through convolutional networks, which necessitate intricate image preprocessing. However, the availability of annotated retinal images with lesion labels remains limited, while extensive manual adjustments are often required to achieve precise lesion detection [8].

The advent of Deep Learning (DL) has brought robust solutions to the medical domain [9] and [10]. The study [11] extracts the green channel from fundus images and enhances DR detection by employing a modified AlexNet with a larger input size and optimized layers, achieving 96.6% for different stages of DR severity. Innovations like Multi-Cell architecture [12] further streamline computation and address gradient vanishing issues. A coarse-to-fine network in [13] is designed to categorize five stages of DR severity, addressing data imbalance issues. The network's attention gate module suppresses irrelevant physiological structures in fundus images, emphasizing lesion areas and extracting discriminative features for accurate DR grading. These structural variations in convolutional networks underscore the importance of reproducibility as the field progresses; therefore, utilizing available pre-trained models, such as through transfer learning, can facilitate broader application. Transfer Learning (TL) has indeed proven useful in DR screening , especially for

U. Phan, O. Devecioglu and M. Gabbouj are with the Department of Computing Sciences, Tampere University, Tampere, Finland (e-mail: uyen.phan@tuni.fi , ozer.devecioglu@tuni.fi, moncef.gabbouj@tuni.fi,).

S. Kiranyaz is with the Electrical Engineering Department, Qatar University, Doha, Qatar (e-mail: mkiranyaz@qu.edu.qa

overcoming data shortages and integrating clinical lesion features [14]-[17]. Image processing techniques enhance model generalization, including augmentations such as flipping, brightness changes, contrast adjustments, and rotations [18]. However, Mohsin et al. [14] applied TL by freezing layers of both GoogleNet and ResNet-18, combining features into a hybrid vector for binary and multi-class classification of fundus images. In [17] an Extreme Gradient Boosting (XGBoost) based DR-detector, trained with features from pre-trained ResNet-50 and image-level clinical features, achieved classification accuracy of 100%—a statistically exceptional yet improbable result given the limited test cohort (n=162, with only 81 DR-positive cases). While TL approaches perform well with such very small datasets, they require high computational complexity, making them impractical for large datasets and real-time systems. Recent advancements include CANet [19], featuring a disease-specific attention module for extracting individual disease features and a disease-dependent attention module for understanding inter-disease relationships. A convolutional Block Attention Module (CBAM) for DR severity detection from single Color Fundus Photographs (CFP) was proposed in [20].

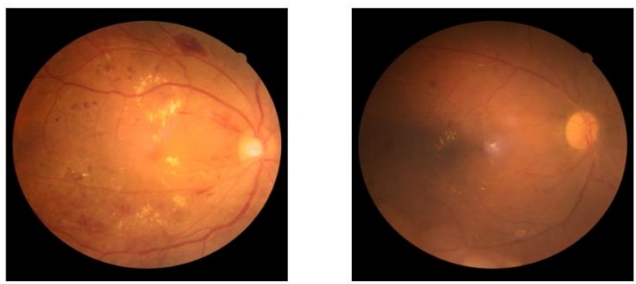

**Figure 1** High (left) and low (right) quality fundus images with the same diabetic retinopathy level. The lack of contrast, blurring especially on the arteries, and black patches are noticeable on the right image.

Despite extensive research on detecting diabetic retinopathy (DR) using fundus images, the accuracy of pathological detection heavily depends on high-quality imaging. Figure 1 presents two fundus images of the retina, both depicting the same DR severity level but significantly differing in quality. The high-quality image on the left is sharp and well-illuminated, clearly displaying retinal features such as the vascular arch and lesions. In contrast, the low-quality image on the right exhibits poor lighting, shadowing over the posterior pole, and blurriness, which obscure crucial details and hinder accurate DR grading. This disparity highlights the need to enhance degraded images, especially by a blend of artifacts, to ensure consistent and reliable evaluation. To address this challenge, we propose a novel Progressive TL (PTL) based multi-pass image restoration method designed to improve the overall quality of fundus images progressively. Traditional image-to-image restoration approaches primarily focus on learning a direct transformation between the pairs of low-quality and high-quality images. However, these methods may struggle to fully recover degraded images with a random blend of artifacts, thus leaving behind residual artifacts or incomplete details. To overcome this limitation, the proposed approach introduces a multi-pass and blind refinement process without requiring such paired low- and high-quality images. First, after training the model on a set of low-quality and high-quality images, the model restores the low-quality samples in the training set. The restored low-quality images in the first pass are now used as the new "low-quality" set to train the model in the second pass. By comparing the restored outputs to their high-quality counterparts, the model identifies and applies the necessary corrections to further enhance the required image clarity and details. Moreover, in the second pass, to conduct the restoration ability learned during the first training pass, the best model parameters are used to initiate the model training in the second pass. Therefore, instead of initializing the network randomly for each pass, we leverage the best-performing model from the previous step, train it over the newly formed data and thus accumulate the learned knowledge in each pass. Such a *progressive* transfer learning (PTL) strategy enables the model to not only perform an initial conversion but also iteratively refine images, leading to superior restoration results. In the following passes, this practice is repeated, and thus at the end, a cascaded set of restoration networks can be used to iteratively restore the low-quality images.

As the restoration model, we utilize CycleGAN [21], a state-of-the-art image-to-image transformation model, which can learn the blind restoration over the unpaired image sets. CycleGAN aims to generate outputs that closely resemble the ground truth during training, but it can sometimes misclassify distorted images as high-quality ones due to its adversarial training mechanism. To surpass this limitation, we integrate PTL into its training and thus, enhancing its restoration potential through a multi-pass refinement process. By leveraging such progressive learning, our method builds upon CycleGAN's strengths while mitigating its shortcomings, ultimately achieving more reliable and detailed fundus image restoration.

Overall, the novel and significant contributions can be summarized as follows:

- We propose the first blind restoration approach for fundus images using Cycle-GANs.
- A pioneer multi-pass restoration scheme is proposed to iteratively restore the low-quality images by a cascaded set of restoration networks.
- Each model used in each restoration pass is trained using the proposed PTL in order to accumulate the learned restoration abilities via the proposed multi-pass training which in turn boosts the overall restoration performance.
- The proposed novel approach is evaluated both qualitatively and quantitatively. Especially, the quantitative evaluations reveal that the DR diagnosis accuracy can significantly be improved at each pass.
- Finally, this pilot study demonstrates that even the high-quality images selected by the experts can further be improved and thus can be used as new ground-truth high-quality fundus images.

The rest of this article is organized as follows: The proposed multi-pass training with the PTL is presented in Section II with a brief explanation of CycleGANs. In Section III, the benchmark DeepDRID dataset will first be introduced. and then a detailed set of both qualitative and quantitative evaluations will be presented. Conclusions and topics for future research will be discussed in Section IV.

## II. METHODOLOGY

In this section, we first provide a brief overview of unpaired image translation model, Cycle-GANs and their key characteristics. Next, the proposed multi-pass training approach with the proposed PTL will be introduced.

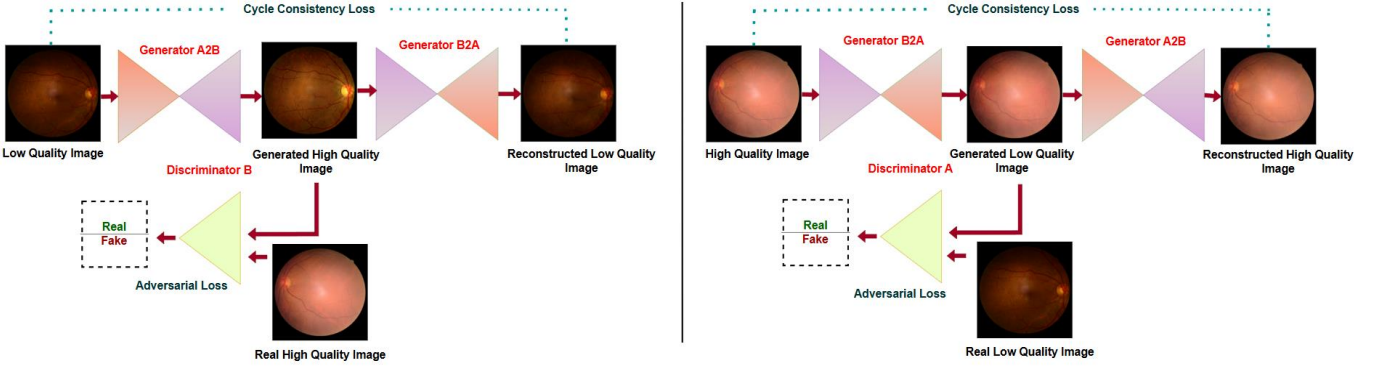

**Figure 2 General Framework of CycleGAN for Fundus Image Quality Restoration.** Starting from the left side, Generator A2B converts a low-quality fundus image into a high-quality counterpart. Discriminator B assesses whether the generated image is real or fake, imposing adversarial loss $L_{adv}$ against real high-quality images. Generator B2A then reconstructs the original low-quality image, maintaining cycle consistency $L_{cyc}$. The reverse process occurs on the right panel, where Generator B2A maps a high-quality image to a low-quality version, and Discriminator A distinguishes real low-quality images from generated ones.

### A. Cycle Consistent Adversarial Networks (Cycle-GANs)

CycleGAN [21] is a GAN model that enables bidirectional image translation between two domains without requiring paired data, making it especially valuable for fundus image restoration, where acquiring corresponding low- and high-quality image pairs is difficult. Figure 2 shows the general framework of the Cycle-GAN. The model consists of two generators and two discriminators. One generator converts low-quality fundus images $X_L$ to high-quality $X_H$, while the other performs the reverse transformation, ensuring cycle consistency. Meanwhile, the discriminators distinguish real and generated (fake) images, pushing the generators to produce visually convincing outputs while preserving domain-specific features. The most crucial component of the CycleGAN training is the cycle consistency loss $L_{cyc}$ which ensures images can be reconstructed after transformation, preserving visual fidelity and pixel integrity. The $L_{cyc}$, is formulated as in Eq. (1)

$$L_{cyc}(G_{L \to H}, G_{H \to L}, X_L, X_H)$$
$$= \frac{1}{m}\sum_{i=1}^{m}\left[|G_{H \to L}\left(G_{L \to H}(X_L(i))\right) - X_L(i)|_1\right]$$
$$+ \frac{1}{m}\sum_{i=1}^{m}\left[|G_{L \to H}\left(G_{H \to L}(X_H(i))\right) - X_H(i)|_1\right] \quad (1)$$

Where the generator $G_{L \to H}$ refines the low-resolution fundus images while maintaining critical spatial, morphological, and pathological features. Preserving diagnostic markers like microaneurysms and vascular distortions is crucial for reliable diagnosis, regardless of visual clarity improvements. To ensure authenticity, discriminators differentiate real and generated images using adversarial loss $L_{adv}$. By penalizing deviations from high-quality data, they guide generators to produce realistic outputs. A higher adversarial loss indicates improved generator performance, aligning with Eq. (2).

$$L_{adv1}(G_{L \to H}, D_H, X_L) = \frac{1}{m}\sum_{i=1}^{m}\left(1 - D_H\left(G_{L \to H}(X_L(i))\right)^2\right)$$
$$L_{adv2}(G_{H \to L}, D_L, X_H) = \frac{1}{m}\sum_{i=1}^{m}\left(1 - D_L\left(G_{H \to L}(X_H(i))\right)^2\right) \quad (2)$$

Here, $D_H$ distinguishes real high-quality images from generated ones, while $D_L$ does the same for the low-quality domain.

Alongside adversarial and cycle consistency losses, identity loss $L_{ide}$ ensures that images already in the target domain remain unchanged, preventing unnecessary modifications that could degrade quality.

$$L_{ide}(G_{L \to H}, G_{H \to L}, X_L, X_H)$$
$$= \frac{1}{m}\sum_{i=1}^{m}\left[|G_{L \to H}(X_H(i)) - X_H(i)|_1\right]$$
$$+ \frac{1}{m}\sum_{i=1}^{m}\left[|G_{H \to L}(X_L(i)) - X_L(i)|_1\right] \quad (3)$$

The objective of training the CycleGAN model is to minimize the cost function, as expressed by the summation dynamic in Eq. (4).

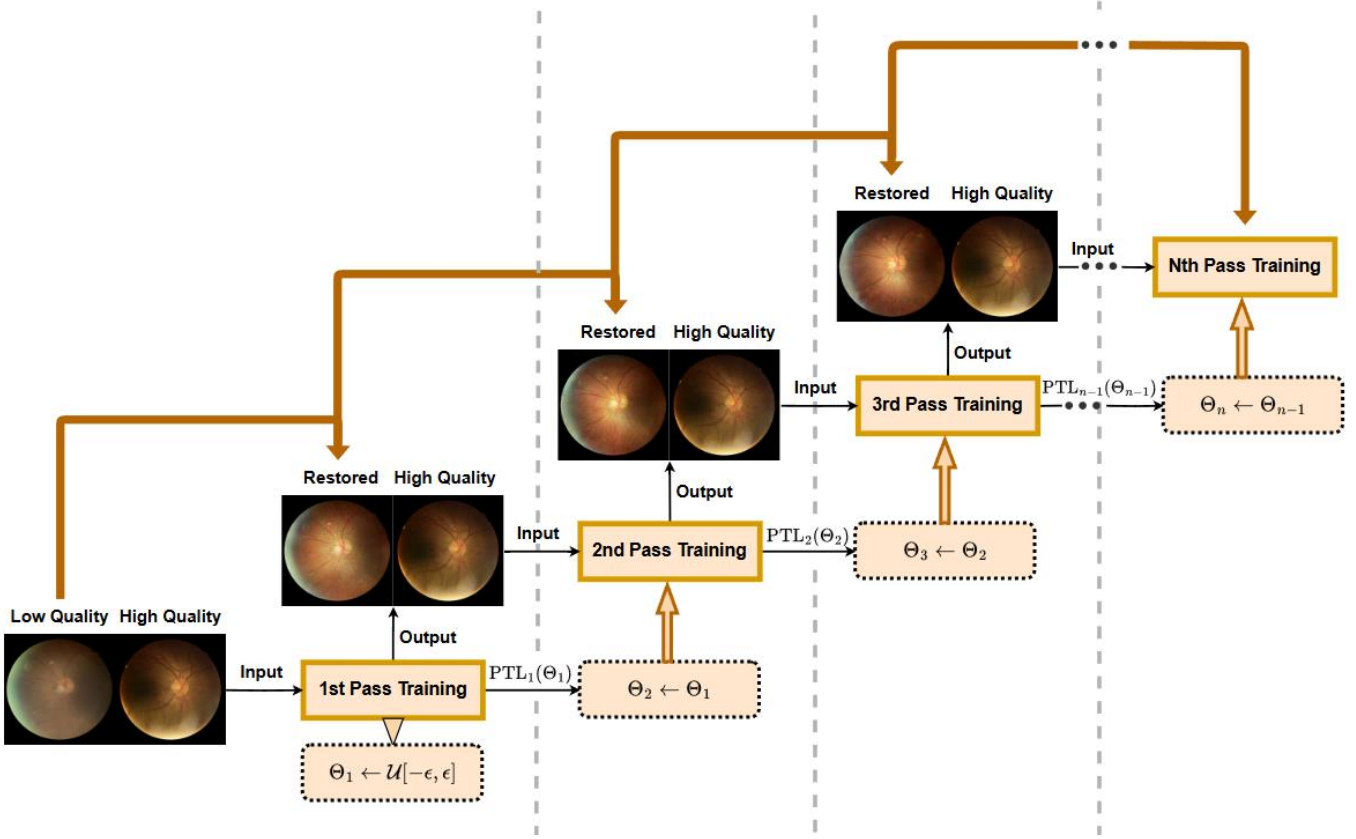

**Figure 3 Progressive Transfer Learning Framework for Retinal Image Restoration.** Initially, the model undergoes 1st Pass Training with parameters $\theta_1 \sim \mathcal{U}[-\epsilon, \epsilon]$ producing a restored image. Each subsequent $n^{th}$ Pass ($n \leq N$) refines the model using parameters $\theta_n \leftarrow \theta_{n-1}$ and progressively improved inputs, denoted as $PTL_n(\theta_{n-1})$. After $N$ training passes, the final model $\theta_N$ achieves the best high-quality image restoration through iterative improvements.

$$L_{total} = L_{adv1} + L_{adv2} + \lambda L_{cyc} + \beta L_{ide} \quad (4)$$

To balance the influence of the discriminators, adaptive weights λ and β are specified as hyperparameters to control the relative weights of the cycle-consistency and identity losses, respectively.

### B. Progressive Transfer Learning Approach

While Cycle-GAN excels at transforming low quality images to high quality counterparts, its single-pass restoration approach can sometimes fail to fully restore over multi-artifact-corrupted images, as the first restoration may even leave certain artifacts unrestored. In this section, the novel Progressive Transfer Learning (PTL) approach that is embedded into the proposed multi-pass training is introduced. Therefore, the resultant multi-pass restoration approach via cascaded restoration networks can address the challenges of iteratively restoring low-quality images in each pass to maximize their quality. Figure 3 illustrates how the PTL takes advantage of knowledge cumulated from the previous passes as a springboard for progressive learning, resolving the challenges of catastrophic forgetting [23]. Initially, the CycleGAN model is trained with random initialization for image-to-image translation between low-quality and high-quality images. The restored output of low-quality images from this initial pass are then used as inputs for the next training pass. The first PTL is applied on both generator and discriminator networks that are initialized with the best network parameters achieved in the first training pass. Such a PTL procedure is repeated at each pass to enable the fine-tuning to progressively refine the outputs. As a result, this iterative process ensures that prior knowledge and latent feature spaces learned in earlier stages are capitalized upon, allowing the model to achieve a superior restoration at each pass. Furthermore, the results demonstrate that by transferring learned features across few training passes, PTL facilitates the recovery of fine-grained details and improves alignment with high-quality ground truth images. Such a multi-pass restoration approach with PTL also eliminates the need for any prior training over other datasets while fully utilizing the prior knowledge obtained via pre-trained network parameters, making PTL an efficient and powerful solution for any iterative image restoration.

The initial training pass begins with the original DeepDRiD dataset $D(0) = \{X_L^{(0)}, X_H\}$, comprising low-quality images $X_L^{(0)}$ and their high-quality counterparts $X_H$, as shown in Figure 1. In this pass, a base CycleGAN model $M$ is trained, with the model weights initialized *randomly* within $\mathcal{U}[-\epsilon, \epsilon]$. The best-performing generator and discriminator networks are selected at epoch $\Theta$ in which discriminator loss $L_{adv1}$ is maximized on the validation set — a choice that will be justified in Section III. $Q$ denotes the weights stored in what is referred to as the latent space as explained above in the PTL context.

$$Q(1) = \arg \max_{Q(1)} L_{\text{adv1}}^{(1)} \qquad (5)$$

This approach enables the evaluation of the realism of the outputs for subsequent restoration. The preferred model is then applied to the low-quality images to produce an intermediate training dataset with the restored images, $X_L^{(1)}$.

$$X_L^{(1)} = M_{Q(1)}(X_L^{(0)}) \qquad (6)$$

In subsequent training passes, $i = 2, 3, \ldots, n$, the low-quality image set is updated for the datasets, $D(i) = \{X_L^{(i)}, X_H\}$. The next selection of the best epoch $\Theta_n$ is again guided by $\max_{Q(i)} L_{\text{adv1}}^{(i)}$.

$$Q(i) = \arg \max_{Q(i)} L_{\text{adv1}}^{(i)} \qquad (7)$$

where the subsequent dataset refinement practice remains intact, i.e.,

$$X_L^{(i+1)} = M_{Q(i)}(X_L^{(i)}) \qquad (8)$$

The final restored output after *n*-restoration passes, $X_L^{(n)}$ can be obtained by sequentially using the best restoration models from all passes and can be expressed as,

$$X_L^{(n)} = M_{Q(n)}\left[M_{Q(n-1)}\left[\ldots M_{Q(1)}[X_L^{(0)}]\right]\right] \qquad (9)$$

## III. EXPERIMENTAL RESULTS

This section will first outline the details of the DeepDRiD dataset, followed by the preprocessing steps applied to the set. Next, a detailed set of qualitative and quantitative evaluations performed over this benchmark dataset will be detailed and the results will be discussed.

### A. DeepDRID Dataset

The experiments in this study were conducted over the DeepDRiD dataset [24], which was obtained from the ISBI-2020 Challenge. This dataset now serves as a benchmark for evaluating DR detection models. It includes a comprehensive set of fundus images designed to support DR gradings, which is assessed across several criteria, including the presence of artifacts, image clarity, field definition, and an overall quality score. Each image is labeled as either "low quality" (0) or "high quality" (1) based on its overall quality. In terms of image quality, the dataset consists of 842 low-quality images and 758 high-quality images. Regarding the presence of diabetic retinopathy, the dataset includes 480 and 1120 abnormal cases.

### B. Experimental Setup

The architectures of the generator and discriminator networks used in this study are illustrated in Figure 4. For all experiments Cycle-GAN's U-Net generator comprises 7 layers for down sampling and corresponding 7 layers for up sampling, interconnected through skip connections. Each processing block employs max pooling with a 2×2 kernel to reduce the spatial dimensions by half. During the down sampling phase, the number of filters increases while the spatial dimensions decrease, and vice versa during up sampling. Both the convolutional and transposed convolutional layers utilize a kernel size of 4 and a stride of 2. Padding is set to 1 to ensure proper alignment and avoid any dimensional mismatches in the output. In this setup, the PatchGAN discriminator features 5 2D convolutional layers, each with a kernel size of 4. Between these layers, batch normalization and Leaky ReLU activations are applied. The final layer produces a single-channel output with a spatial resolution of 30×30. Each element in the grid reflects the discriminator's authentication of whether the corresponding patch of the input fundus image is of high or low quality.

The experiment follows a training scheme with a maximum of 3500 backpropagation epochs. The initial learning rate is set to $2x10^{-4}$ for the first 100 epochs, after which we applied a decay rate of $10^{-5}$ for another 100 epochs to both the generators and discriminators. This fine-tuning practice establishes itself as an indispensable procedure in the Progressive Transfer Learning (PTL), facilitating the model shifts from rapid learning to more incremental adjustments. During subsequent PTL applications, these hyperparameters are fine-tuned to optimize the model's performance.

The training dataset comprises 3,500 low-quality images, with 868 images allocated for validation, while the test set consists of 218 high-quality images. To enhance CycleGAN's generalizability, we adopt a systematic augmentation strategy in dataset splitting, reducing the risk of overfitting to specific patterns. As highlighted in Section II, one of our key objectives is to maximize the adversarial loss of the discriminator, prioritizing training epochs where the discriminator loss remains high. This approach reinforces the adversarial dynamic in CycleGAN, where the generator and discriminator engage in a min-max optimization process, ultimately leading to more effective image restoration.

The qualitative evaluations reveal that the proposed multi-pass restoration improves both the quality and the clarity over the important details on fundus images, which in turn will obviously improve the diagnosis accuracy. In order to measure the improvement quantitatively and demonstrate the importance of restoration on the DR diagnosis, identical CNN classifiers are trained, one over the original and the other three over the restored dataset by each pass. By comparing their DR diagnosis performances on these datasets, we gauge the degree of restoration achieved by each restoration pass and its impact on downstream classification tasks. To ensure a reliable and insightful comparison, we used the standard performance metrics commonly used in the literature [26] such as Accuracy, Precision, Sensitivity (Recall), and F1 score.

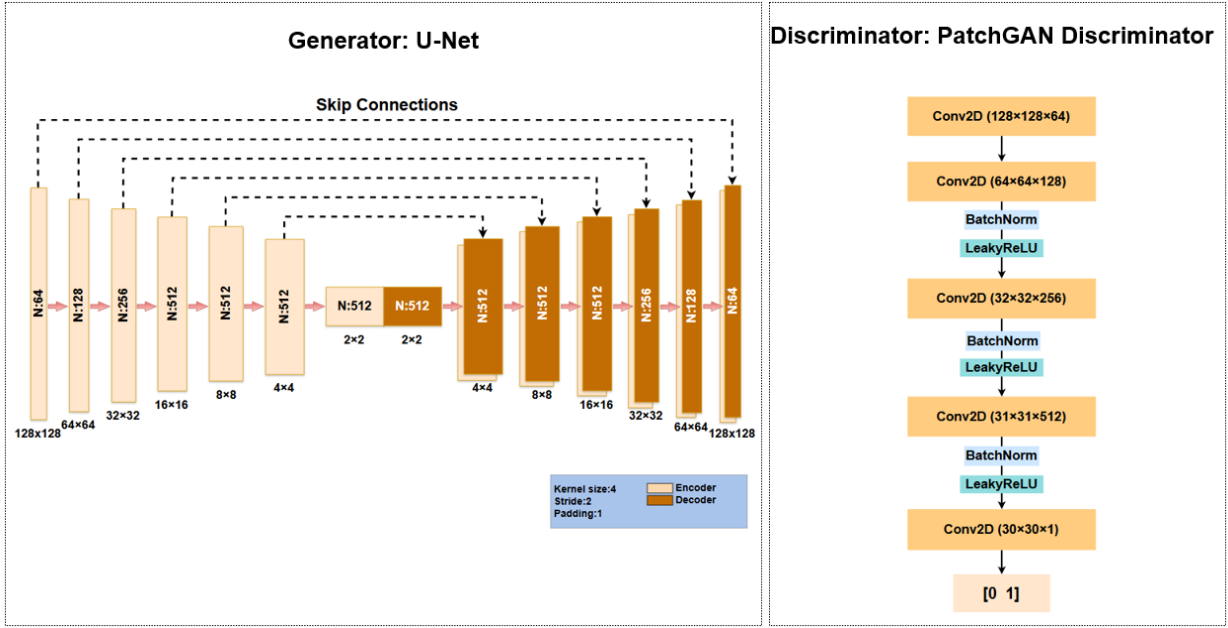

**Figure 4** The Generator and Discriminator Architecture of the CycleGAN Model.

We prioritize epochs with high discriminator loss, as stronger adversarial feedback ensures meaningful restoration in CycleGAN. Early epochs are not favored since the discriminator is weak and easily fooled by visually plausible but clinically inadequate outputs. After 100 epochs, meaningful restoration begins to occur when low generator loss aligns with high discriminator loss, preserving critical retinal details. After restoring the low-quality images using our proposed method, we employed a compact CNN classifier for the quantitative evaluations. This classifier features two main components: a feature extraction module and a classification head, as shown in Figure 5, The feature extraction module includes three convolutional layers, each followed by a ReLU activation function and a max-pooling operation. Then the classification processes the extracted features by flattening them and passing them through two fully connected layers, with ReLU activation applied for non-linearity. The final output layer generates two logits, corresponding to the classification labels—normal or abnormal.

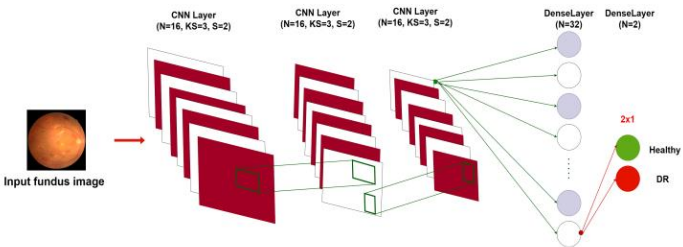

**Figure 5** The CNN classifier architecture.

### C. Qualitative and Quantitative Evaluations

For the qualitative evaluations, Figure 6 shows the input low-quality input fundus images in the test set and the 3-pass restoration results along with a similar high-quality reference image. It is apparent from the figure that the low-quality inputs suffer severely from blurriness, dark patches, saturation and lack of contrast. The first pass restoration results show certain improvements; however, such artifacts are still visible. Each restoration pass afterwards gradually diminishes them and progressively refines the images by enhancing sharpness, correcting colors, and improving contrast at each pass. More restoration results are shown in Appendix A.

Figure 7 shows some sample 3-pass restoration results where the networks are trained with and without PTL (i.e., with random initialization). The results clearly indicate the substantial role of PTL on the multi-pass restoration performance. When PTL is not used and networks are trained with random initialization as in the first pass, then the quality improvement obtained in each pass is limited and sometimes can even get worse.

Figure 8 illustrates two sample restoration results from the test set where dark-patch removals are highlighted with a red rectangle. Moreover, as in the other results in Figure 6, the improvement in the important details such as vascular structures and lesion regions can be better perceived at the last restoration pass (3$^{rd}$ pass). Finaly, the proposed multi-pass restoration approach with PTL can even improve the ground-truth high-quality images where some sample restoration outputs are shown in Figure 9. Such ground-truth set of high-quality images may still contain subtle artifacts due to the complementary nature of image pairs, where one image may be degraded while its counterpart remains suitable for grading. This is quite clear, especially at the top HQ image, which suffers from both dark and white patches and the 3$^{rd}$ restoration pass significantly removes both. At the bottom HQ image, the dark patches are significantly reduced and more importantly, the blurred vascular structure is improved as well.

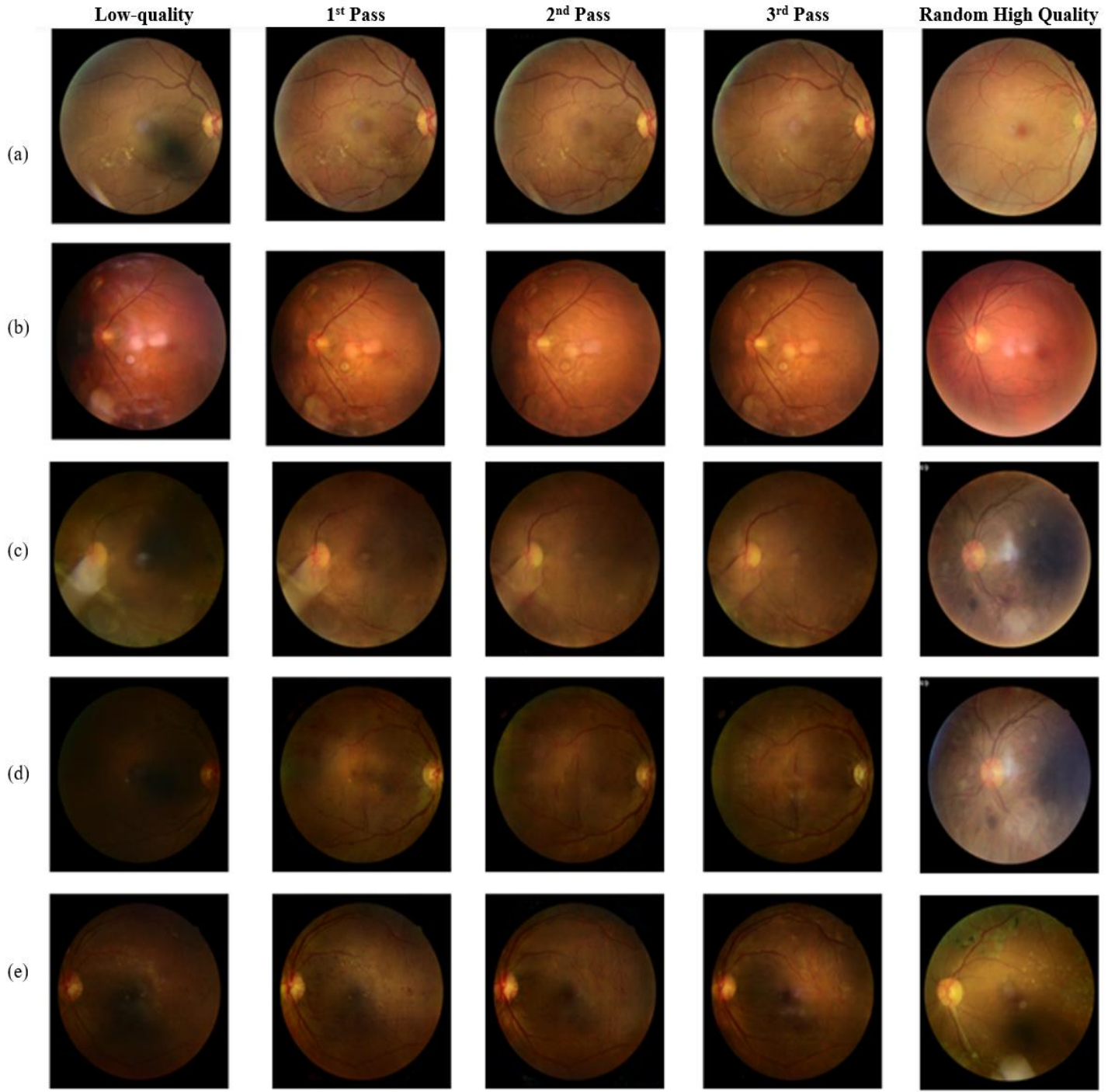

**Figure 6** 3-pass sample fundus image restoration results.

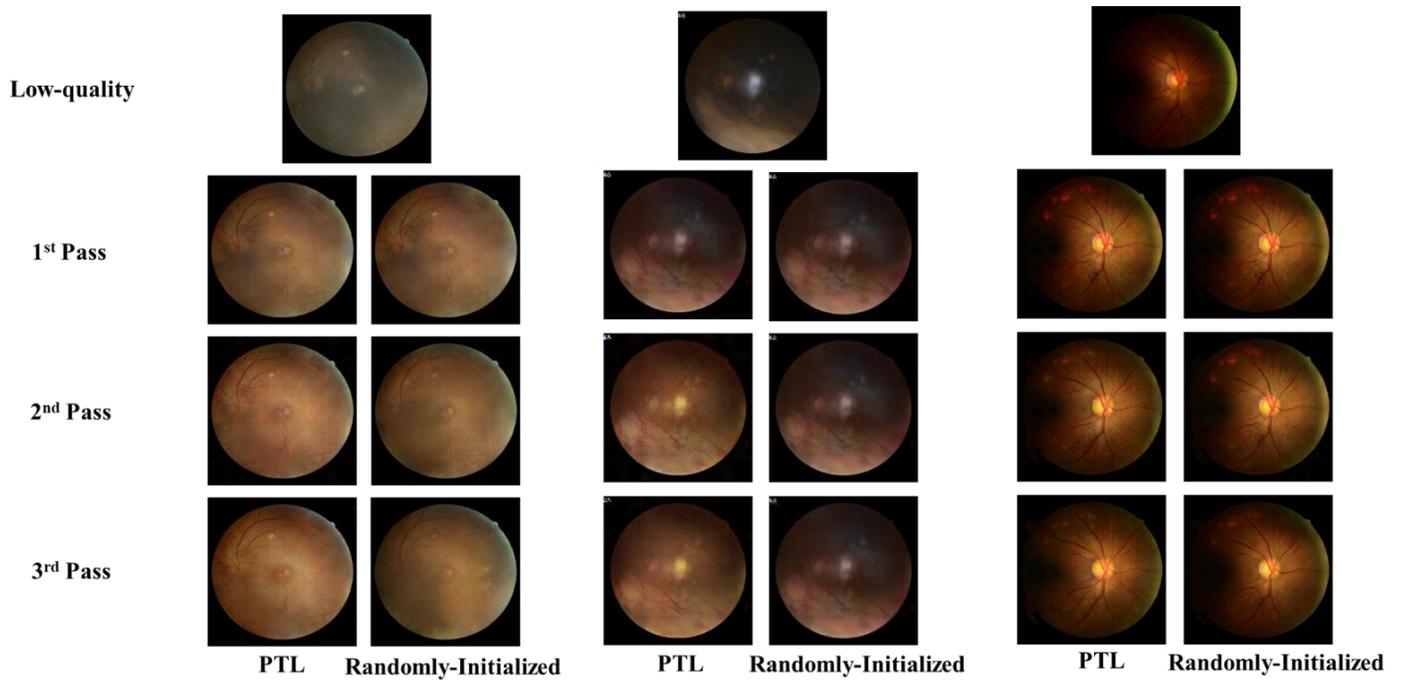

**Figure 7** Sample 3-pass restoration results where models are trained with random initialization and PTL.

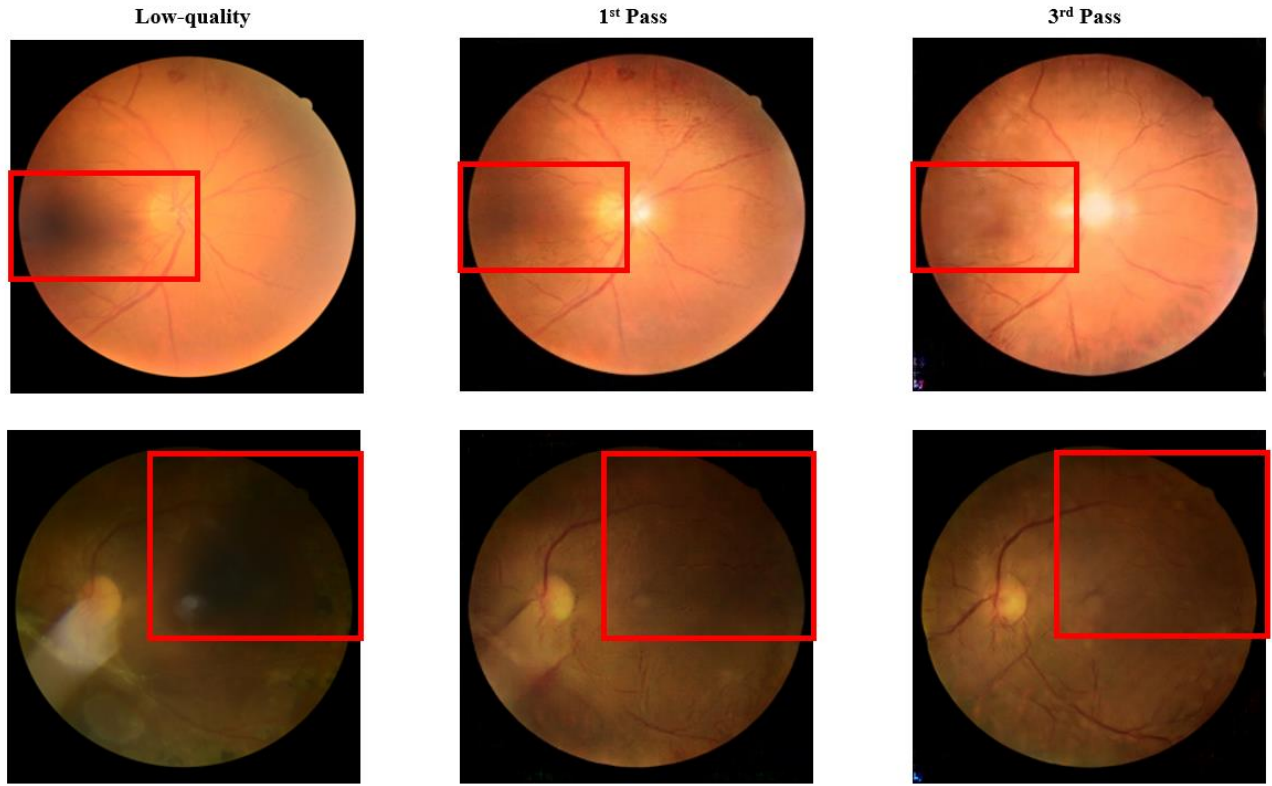

**Figure 8** Sample multi-pass restoration results where dark-patch removals are highlighted with a red rectangle.

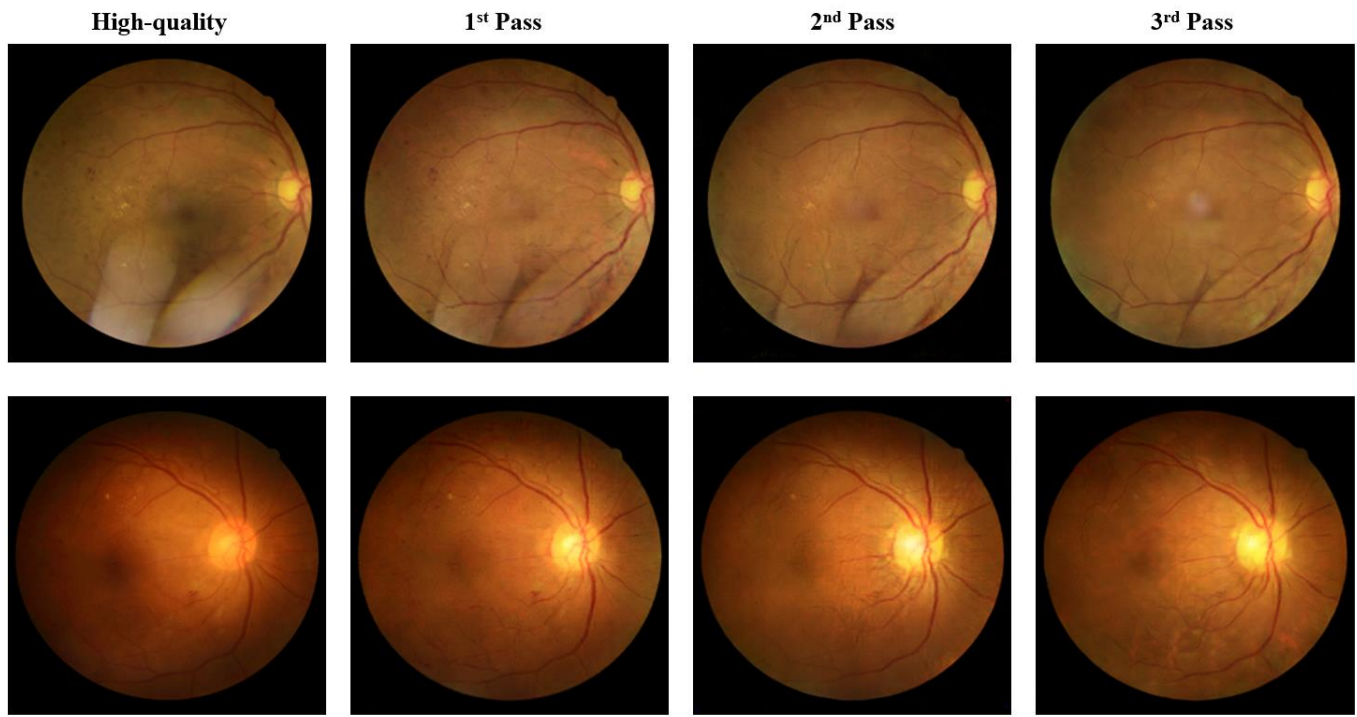

**Figure 9** Sample fundus image restoration results from high quality images.

**Table 1** Quantitative evaluations for DR detection in terms of Accuracy, Sensitivity, Precision and F1 Score.

| Dataset | Accuracy | Precision | Sensitivity | F1 Score |
|---|---|---|---|---|
| Original | 88.00 | 81.03 | 78.33 | 79.66 |
| 1st Pass Restoration | 90.00 | 78.57 | 91.67 | 84.62 |
| 2nd Pass Restoration | 91.00 | 79.58 | 94.17 | 86.26 |
| **3rd Pass Restoration** | **93.50** | **83.57** | **97.50** | **90.00** |

The diagnosis performance results presented in Table 1 demonstrate the significant improvements in the classification performance achieved by each consecutive restoration pass. The highest classification performance in all metrics is achieved at the last (3rd) restoration pass. Sensitivity (Recall), especially, is the most important metric that indicates classifier's ability to detect the DR. With the proposed restoration approach, the overall improvement on the Sensitivity is above 19%. Finally, the harmonic means of Precision and Recall, F1 is improved more than 10%, which is substantial.

### D. Computational Complexity Analysis

The network size, total number of parameters (PARs), and inference time for each network configuration are computed in this section. The Narvi SLURM cluster at Tampere University was used for all experiments. The cluster is accessible via the frontends narvi.cc.tut.fi and narvi-shell.cc.tut.fi, with each CPU node featuring 24-core E5-2620 processors and 128GB of memory. The experiment utilized networks with the following configurations: G_A and G_B each had 222.063M parameters, while D_A and D_B had 8.292M parameters. Processing a single fundus image on a single CPU takes about 2.8 seconds. This is our key limitation, indicating that further optimization or more powerful hardware is needed for real-time processing.

## IV.  CONCLUSION

The increasing global prevalence of diabetic retinopathy underscores the critical need for early detection and effective intervention. High-quality fundus imaging, combined with advanced machine learning techniques, has proven to be invaluable in improving diagnostic accuracy. However, challenges such as poor-quality images remain prevalent, especially in resource-constrained settings. This study introduces a novel PTL-based multi-pass restoration method that enhances low-quality fundus images, making them suitable for a more accurate DR diagnosis. The proposed restoration approach is especially effective at preserving fine vascular structures, lesion boundaries, and pathological markers. By employing the proposed multi-pass training with PTL, each restoration pass benefits from the accumulated knowledge and thus can address the limitations of traditional image-to-image restoration methods such as over-smoothing or saturation, thus ensuring that critical features are preserved. The experimental results demonstrate the effectiveness of our method in enhancing the quality of fundus images progressively, which in turn improves the performance of DR classification models significantly. Overall, the proposed approach offers a promising solution to the challenges posed by blind image restoration in general, especially when the images are corrupted by a random blend of artifacts with varying severities. For future work, it would be interesting to see how PTL-based multi-pass restoration, acting as a non-ad hoc solution, could generalize to other medical imaging modalities. For example, exploring its application in X-ray, MRI and CT image restoration, where progressive refinement is essential for optimal diagnosis, will be the topics of our future work.

APPENDIX A

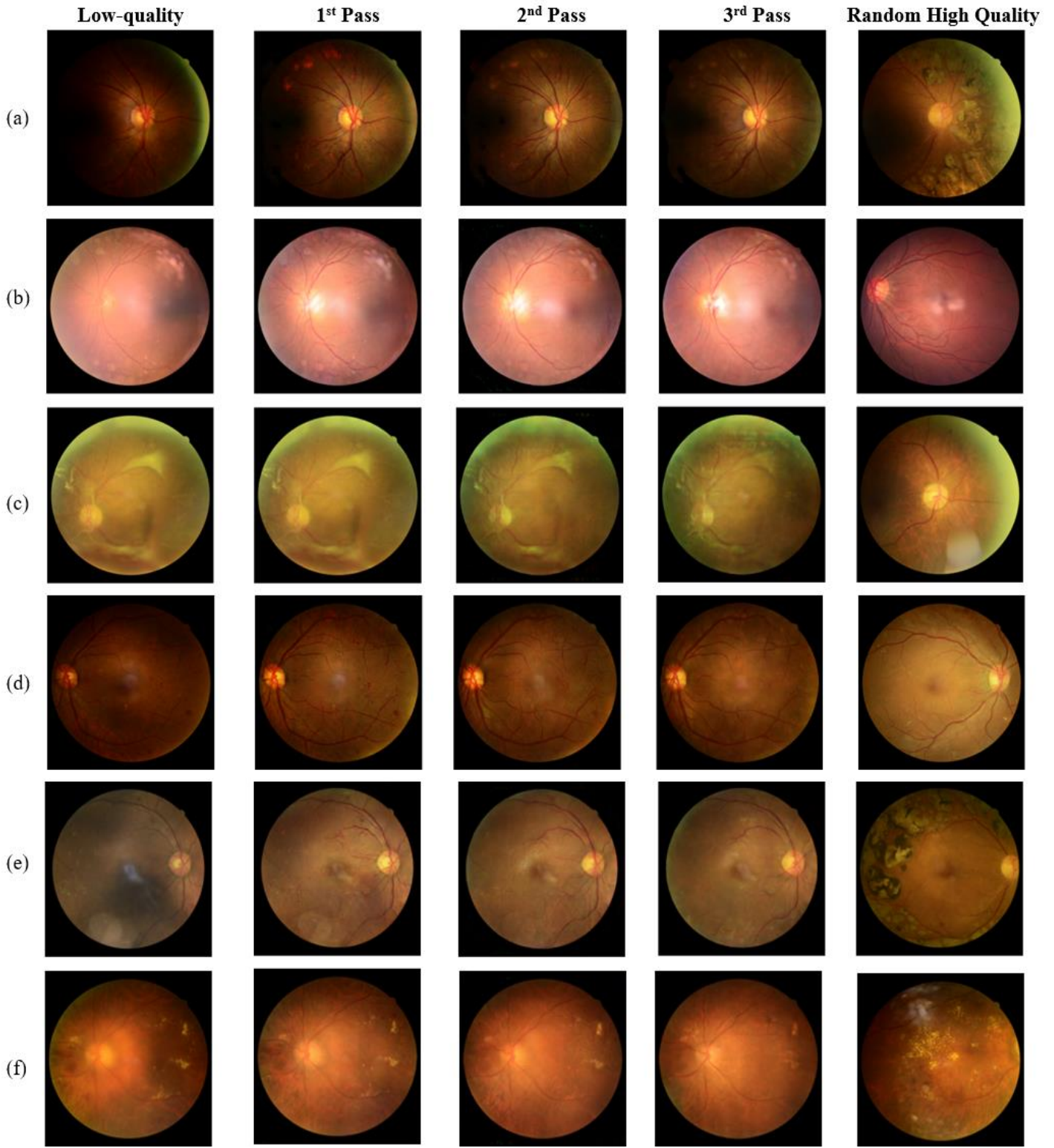

**Figure 10** Supplementary 3-pass sample fundus image restoration results.

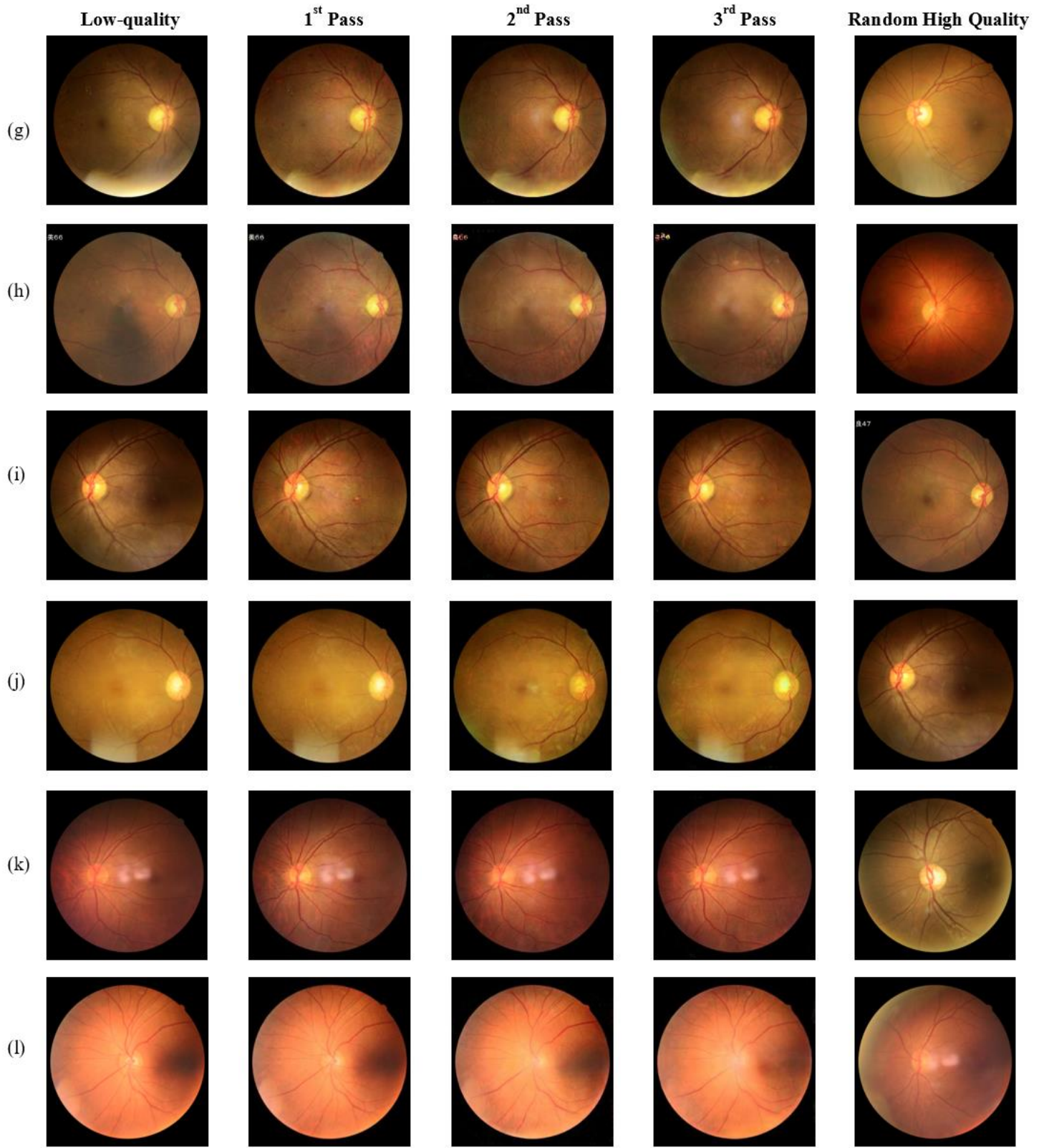

**Figure 11** Supplementary 3-pass sample fundus image restoration results.

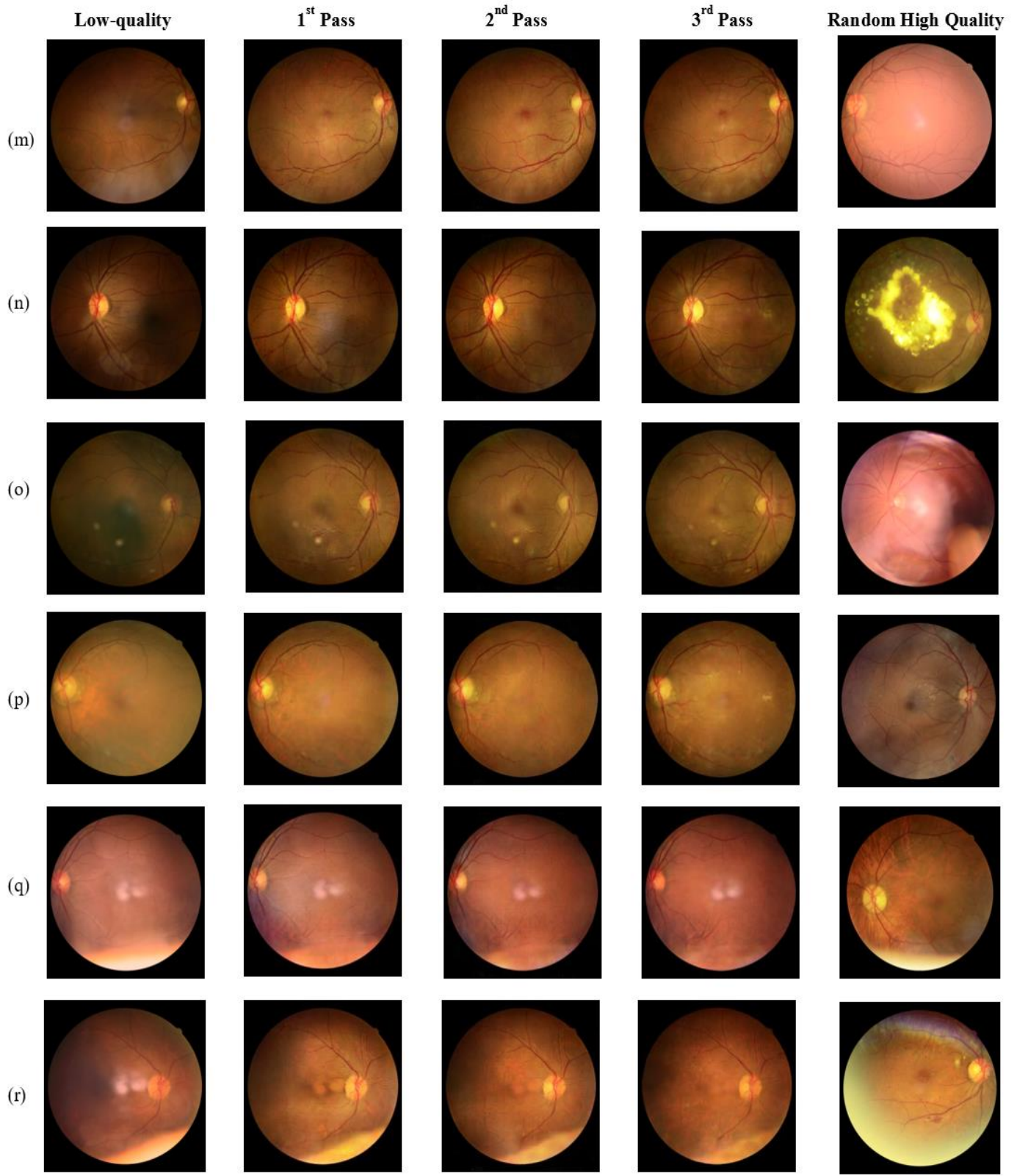

**Figure 12** Supplementary 3-pass sample fundus image restoration results.